Rayleigh-Taylor instability with variable acceleration

Des L. Hill (1); Aklant K. Bhowmick (2); Snezhana I. Abarzhi* (1)

The University of Western Australia, AUS (1); Carnegie Mellon University, USA (2)

*corresponding author: snezhana.abarzhi@gmail.com

We consider the long-standing problem of Rayleigh-Taylor instability with variable acceleration, and focus on the early-time dynamics of an interface separating incompressible ideal fluids of different densities subject to an acceleration being a power-law function of time for a spatially extended three-dimensional flow periodic in the plane normal to the acceleration with symmetry group p6mm. By employing group theory and scaling analysis, we discover two distinct sub-regimes of the early time dynamics depending on the exponent of the acceleration power-law. The time-scale and the early-time dynamics are set by the acceleration for exponents greater than -2, and by the initial growth-rate (due to, e.g., initial conditions) for exponents smaller than -2. At the exponent value (-2) a transition occurs from one regime to the other with varying acceleration strength. For a broad range of the acceleration parameters, the instability growth-rate is explicitly found, the dependence of the dynamics on the initial conditions is investigated, and theory benchmarks are elaborated.

## Section 1 - Introduction

Rayleigh-Taylor instability (RTI) controls a broad range of processes in nature and technology, in fluids, plasmas and materials [1,2]. RTI develops when the fluids of different densities are accelerated against their density gradients [3,4]. Intense interfacial Rayleigh-Taylor (RT) mixing of the fluids ensues with time [1,2]. Examples of RT-relevant processes include inertial confinement fusion, supernova explosion, material transformation under impact, as well as fossil fuel recovery and nano-electronics [1,2]. In realistic environments RTI is often driven by variable acceleration, whereas the bulk of existing studies is focused on the cases of sustained and impulsive accelerations [1,2]. The case of impulsive acceleration induced by a shock is referred as Richtmyer-Meshkov (RM) instability [5,6]. In this work we consider the long-standing problem of RTI subject to an acceleration being a power-law function of time [1,2,7]. We focus on the early-time dynamics, and, by applying group theory and scaling analysis [7-9], identify the instability growth-rate for a broad range of acceleration parameters and initial conditions. We find two sub-regimes depending on the acceleration exponent - acceleration-driven RT-type and initial-growth-rate-driven RM-type, with each sub-regime having its own time-scale, type of dynamics, and theory benchmarks.

RT flows, while occurring in vastly different physical circumstances, have some similar features of the evolution [9]. RTI starts to develop when the fluid interface (or, the flow fields) is (are) slightly perturbed near the equilibrium state [3,4]. Over time the interface is transformed to a composition of small-scale shear driven vortical structures and a large-scale coherent structure of bubbles and spikes, with the bubble (spike) being a portion of the light (heavy) fluid penetrating the heavy (light) fluid [8]. Eventually, the flow transits to the stage of intensive interfacial mixing [7].

Non-equilibrium RT dynamics is extremely challenging to study due to, e.g., tight requirements on the flow control and diagnostics in experiments, the need to track interfaces and capture small-scale processes in simulations, the necessity to account for the non-local and singular character of the dynamics in theory [1,2]. Remarkable success has been recently achieved in understanding of the fundamentals of RTI and RT mixing [1,2,7-16]. In particular, the group theory approach has found that nonlinear RTI has a multi-scale character and that RT mixing with constant acceleration may keep order, thus explaining the observations [7-10].

A number of important aspects of RT dynamics still require a deeper understanding. One of them is variable acceleration. RT flows with variable accelerations occur commonly in nature and technology [1,2]. These include RTI induced by unsteady shocks in inertial confinement fusion, blast-wave-driven RT mixing in core-collapse supernovae, RT-unstable plasma irregularities in the Earth's ionosphere, and fossil fuel recovery in industry [17-22]. RTI and RT mixing with variable acceleration is a long-standing problem [1,2]. Only limited information is currently available on RT dynamics under these conditions [22-24].

Here we consider RTI subject to variable acceleration with power-law time dependence. On the side of fundamentals, power-law functions are important to study because they may result in new invariant and scaling properties of the dynamics [25,26]. On the side of applications, power-law functions can be used to adjust the acceleration's time-dependence in realistic environments and thus ensure practicality of our results [17-21]. We consider a three-dimensional spatially extended periodic flow, apply group theory and scaling analysis [7,8], identify the dependence of the instability growth-rate on the acceleration's exponent and strength, and elaborate theory benchmarks for observations. Particularly, we find that the early-time dynamics is set by the acceleration for acceleration exponents greater than -2, and by the initial growth-rate for acceleration exponents smaller than -2, and that at the exponent value (-2) the transition occurs between the sub-regimes by varying the acceleration strength [22-24].

**Section 2 – Governing equations and dynamical system**

The dynamics of ideal fluids is governed by conservation of mass, momentum and energy:

$$\partial\rho/\partial t + \partial\rho v_i/\partial x_i = 0,\ \partial\rho v_i/\partial t + \sum_{j=1}^{3}\partial\rho v_i v_j/\partial x_j + \partial P/\partial x_i = 0,\ \partial E/\partial t + \partial(E+P)v_i/\partial x_i = 0 \quad (1)$$

where $(x_1, x_2, x_3) = (x, y, z)$ are the spatial coordinates with, $t$ is time, $(\rho, \mathbf{v}, P, E)$ are the fields of density $\rho$, velocity $\mathbf{v}$, pressure $P$ and energy $E = \rho(e + \mathbf{v}^2/2)$, $e$ is the specific internal energy [25]. The boundary conditions at the interface are

$$[\mathbf{v} \cdot \mathbf{n}] = 0,\ [P] = 0,\ [\mathbf{v} \cdot \boldsymbol{\tau}] = any,\ [W] = any \quad (2)$$

where $[...]$ denotes the jump of a quantity across the interface; $\mathbf{n}$ and $\boldsymbol{\tau}$ are the normal and tangential unit vectors of the interface with $\mathbf{n} = \nabla\theta/|\nabla\theta|$ and $(\mathbf{n} \cdot \boldsymbol{\tau}) = 0$; $\theta = \theta(x, y, z, t)$ is a local scalar function, with $\theta = 0$ at the interface and with $\theta > 0$ ($\theta < 0$) in the bulk of the heavy (light) fluid that we mark hereafter with sub-script $h(l)$. The specific enthalpy is $W = e + P/\rho$ [25]. The flow is periodic in the plane $(x, y)$ normal to the $z$ direction of the acceleration $\mathbf{g}$, $|\mathbf{g}| = g$ [7,8]. The acceleration is directed from the heavy to the light fluid, $\mathbf{g} = (0, 0, -g)$, and is a power-law functions of time, $g = G t^a$, $t > 0$. Here $a$ is the acceleration exponent, $a \in (-\infty, +\infty)$, and $G$ is the pre-factor, $G > 0$, and their dimensions are $[G] = m/s^{2+a}$ and $[a] = 1$ [22-24]. The flow is free of mass sources

$$\mathbf{v}|_{z \to +\infty} = 0,\ \mathbf{v}|_{z \to -\infty} = 0 \quad (3)$$

The initial conditions are initial perturbations of the flow's fields. We consider incompressible fluids with negligible stratification and density variations. The spatial period (wavelength) $\lambda$ of the initial perturbation sets the dynamics' length-scale [25,27-29].

To solve the problem of RTI with variable acceleration, group theory can be employed [7,8]. By using the techniques of the theory of discrete groups, we apply irreducible representations of a relevant group to expand the flow fields as Fourier series, and further make spatial expansions in a vicinity of a regular point at the interface (i.e., the tip of the bubble or spike). The governing equations are thus reduced to a dynamical system in terms of surface variables and moments, and the solution is sought [7,8,22,23]. We focus on the large-scale coherent dynamics with length scale $\sim \lambda$, presuming that length scale of shear-driven interfacial vortical structures is small, $\ll \lambda$. For convenience, all derivations are performed in the frame of reference moving with velocity $v(t)$ in the $z$-direction, where $v(t)$ is the bubble (spike) tip velocity in the laboratory frame of references [7,8,22,23].

For the large-scale coherent structure, the fluid motion is potential. For symmetry group p6mm, the velocity of the heavy (light) fluid is $\mathbf{v}_{h(l)} = \nabla \Phi_{h(l)}$ with

$$\Phi_h(\boldsymbol{r},z,t) = \sum_{m=0}^{\infty} \Phi_m(t)\left(z+[\exp(-mkz)/(3mk)]\sum_{i=1}^{3}\cos(m\boldsymbol{k}_i\boldsymbol{r})\right)+f_h(t)+cross\ terms, \quad (4)$$

$$\Phi_l(\boldsymbol{r},z,t) = \sum_{m=0}^{\infty} \tilde{\Phi}_m(t)\left(-z+[\exp(mkz)/(3mk)]\sum_{i=1}^{3}\cos(m\boldsymbol{k}_i\boldsymbol{r})\right)+f_l(t)+cross\ terms$$

Here $\boldsymbol{r}=(x,y)$, $\boldsymbol{k}_i$ are the vectors of the reciprocal lattice with $\boldsymbol{k}_i\boldsymbol{a}_j=2\pi\delta_{ij}$, $\boldsymbol{a}_i$ are the spatial periods in the $(x,y)$ plane, $\delta_{ij}$ is the Kronecker delta, with $|\boldsymbol{a}_i|=\lambda$ and $k=|\boldsymbol{k}_i|=4\pi/(\lambda\sqrt{3})$, $\boldsymbol{a}_1+\boldsymbol{a}_2+\boldsymbol{a}_3=0$, $\boldsymbol{k}_1-\boldsymbol{k}_2-\boldsymbol{k}_3=0$, $i,j=1,2,3$, $\Phi_m(\tilde{\Phi}_m)$ are the Fourier amplitudes of the heavy (light) fluid, $f_{h(l)}$ are time-dependent functions, and $m$ is integer [7,8,30].

The fluid interface is $\theta=-z+z^*(\boldsymbol{r},t)$, and function $z^*(\boldsymbol{r},t)$ is defined locally as

$$z^*(\boldsymbol{r},t)=\sum_{N=1}^{\infty}\zeta_N(t)\sum_{i=1}^{3}(\boldsymbol{k}_i\boldsymbol{r})^{2j}+cross\ terms \implies z^*=\sum_{N=1}^{\infty}\zeta_N(t)(x^2+y^2)^{2N}+cross\ terms \quad (5)$$

Here $N$ is natural, $\zeta_N$ are the surface variables, $\zeta_1=\zeta$ is the principal curvature at the bubble (spike) tip. The solution is sought for $t>0$; at late times $\zeta\leq 0, v\geq 0$ for bubbles, and $\zeta\geq 0, v\leq 0$ for spikes.

Upon substituting these expansions in the governing equations and further expanding the equations in the vicinity of the bubble tip, we derive from the governing equations a dynamical system in terms of moments and surface variables. For group p6mm, to first order $N=1$, the interface is $z^*=\zeta(x^2+y^2)$, and the dynamical system is [7,8,24,30]:

$$\dot{M}_0=-\dot{\tilde{M}}_0=-v,\ \rho_h(\dot{\zeta}-2\zeta M_1-M_2/4)=0,\ \rho_l(\dot{\zeta}-2\zeta\tilde{M}_1+\tilde{M}_2/4)=0,$$

$$\rho_h(\dot{M}_1/4+\zeta\dot{M}_0-M_1^2/8+\zeta g)=\rho_l\left(\dot{\tilde{M}}_1/4-\zeta\dot{\tilde{M}}_0-\tilde{M}_1^2/8+\zeta g\right),\ M_1-\tilde{M}_1=any \quad (6)$$

These equations represent absence of sources, continuity of the normal component of velocities and normal component of momentum at the interface, discontinuity of the tangential component of momentum at the interface. Values $M(\tilde{M})$ are the moments, with $M_n=\sum_{m=0}^{\infty}\Phi_n k^n m^n + cross\ terms$ and

$\tilde{M}_n=\sum_{m=0}^{\infty}\tilde{\Phi}_n k^n m^n + cross\ terms$, and integer $n$.

**Section 3 - Results**

   3.1 Early-time dynamics

   For the early-time dynamics only the first order harmonics are retained in the expressions for the momentum, $M_n = k^n \Phi_1$, $\tilde{M}_n = k^n \tilde{\Phi}_1$, $n = 0,1,2$, transforming the dynamical system to:

$$M_0 = -\tilde{M}_0 = -v, \rho_h(\dot{\zeta} - 2\zeta k M_0 - k^2 M_0/4) = 0, \rho_l(\dot{\zeta} - 2\zeta k \tilde{M}_0 + k^2 \tilde{M}_0/4) = 0, \quad (7)$$

$$\rho_h\left(k^2 \dot{M}_0/4 + \zeta \dot{M}_0 - k^2 M_0^2/8 + \zeta g\right) = \rho_l\left(k^2 \dot{\tilde{M}}_0/4 - \zeta \dot{\tilde{M}}_0 - k^2 \tilde{M}_0^2/8 + \zeta g\right), M_0 - \tilde{M}_0 = any$$

The initial conditions at time $t_0$ are the initial curvature $\zeta_0 = \zeta(t_0)$ and velocity $v(t_0)$. The latter sets the initial growth-rate $v_0 = |v(t_0)|$. For a given wavelength $\lambda$ and for $a \neq -2$, there are two time-scales in the dynamics, $\tau_G = (kG)^{-1/(a+2)}$ and $\tau_0 = (kv_0)^{-1}$. At $a = -2$ the time-scale is $\tau_0 = (kv_0)^{-1}$, and the acceleration strength is parameterized by the value $(Gk)$. We consider dynamics for $t_0 > \{\tau_G, \tau_0\}$.

   For a broad class of initial conditions, integration of the governing equations is a challenge [24]. The solution can be found when the initial perturbation amplitude is small and the interface is nearly flat, $|\zeta/k| \ll 1$, particularly, $|\zeta/k| \ll 1/8$ (hence $\ll 1/4$). In this case the system is transformed to

$$\dot{\zeta} = -(k^2/4)v, \dot{v} + (Ak/2)v^2 + (4A/k)\zeta G t^a = 0 \quad (8)$$

Note that the term involving $v^2$ should be kept since it defines the time scale at $g = 0$ [22-24,31].

   Consider the dynamics for $a \neq -2$. With $\sigma = -(\zeta/k), V = v/v_0, T = t/\tau_s$, where $\tau_s$ is some time-scale, the system gets the form

$$d^2\sigma/dT^2 - A\left[\sigma T^a(\tau_s/\tau_G)^{a+2} - 2(d\sigma/dT)^2\right] = 0, d\sigma/dT - \left[(\tau_s/\tau_0)(V/4)\right] = 0 \quad (9)$$

This system indicates the existence of two distinct sub-regimes depending on the interplay of the acceleration parameters and the initial growth-rate – the acceleration-driven and the initial growth-rate driven sub-regimes. Indeed, with $\tau_s = \tau_0$, we find from the first equation in the system

$$\begin{aligned}(\tau_0/\tau_G)^{a+2} \gg 2 &\Leftrightarrow (\tau_0/\tau_G) \gg 2^{1/(a+2)} \Rightarrow (\tau_G/\tau_0) \ll 1 \quad \text{for} \quad a > -2, \\ (\tau_0/\tau_G)^{a+2} \ll 2 &\Leftrightarrow (\tau_0/\tau_G) \ll 2^{1/(a+2)} \Rightarrow (\tau_0/\tau_G) \ll 1 \quad \text{for} \quad a < -2 \end{aligned} \quad (10)$$

   a.  Acceleration exponents greater than -2

   For $a > -2$ the smallest time-scale is $\tau_G$, $\tau_G \ll \tau_0$, the relative contribution of the terms is $|\sigma T^a(\tau_0/\tau_G)^{a+2}| \gg |2(d\sigma/dT)^2|$, and the dynamics is driven by the acceleration. With the time-scale of the fastest process $\tau_s = \tau_G$, the system is transformed to

$$d^2\sigma/dT^2 - A[\sigma T^a] = 0, V - 4(\tau_0/\tau_G)[d\sigma/dT] = 0 \quad (11.1)$$

The solution of the system is

$$\sigma = C_1 \sqrt{T} \, I_{1/2s}\left(\sqrt{A}\, T^s/s\right) + C_2 \sqrt{T} \, I_{-1/2s}\left(\sqrt{A}\, T^s/s\right), \; V = 4(\tau_0/\tau_G)[d\sigma/dT] \quad (11.2)$$

where $s = (a+2)/2$ and $I_p$ is the modified Bessel function of the $p$th order. In complete agreement with the classic results, for constant acceleration, $a = 0$, the solution is transformed to

$$\sigma = C_1 \exp(\sqrt{A}\,T) + C_2(-\sqrt{A}\,T), \; V = 4(\tau_0/\tau_G)[d\sigma/dT] \quad (11.3)$$

In the dimensional form with $\tau_G = (kG)^{-1/(a+2)}$ the solution is

$$(-\zeta/k) = C_1 \sqrt{t/\tau_G} \, I_{1/2s}\left(\sqrt{A}\,(t/\tau_G)^s/s\right) + C_2 \sqrt{t/\tau_G} \, I_{-1/2s}\left(\sqrt{A}\,(t/\tau_G)^s/s\right), \quad (11.3)$$
$$v = (4/k)d(-\zeta/k)/dt$$

b. Acceleration exponents smaller than -2

For $a < -2$ the smallest time-scale is $\tau_0$, $\tau_0 \ll \tau_G$, the relative contribution of the terms is $\left|\sigma T^a (\tau_0/\tau_G)^{a+2}\right| \ll \left|2(d\sigma/dT)^2\right|$, and the dynamics is driven by the initial growth-rate. With the time-scale of the fastest process $\tau_s = \tau_0$, the system is transformed to

$$d^2\sigma/dT^2 + 2A(d\sigma/dT)^2 = 0, \; V - 4(d\sigma/dT) = 0 \quad (12.1)$$

The solution of the system is

$$\sigma = (2A)^{-1} \ln(C_2 T + C_1), \; V = 4(d\sigma/dT) \quad (12.2)$$

or, in the dimensional form

$$(-\zeta/k) = (2A)^{-1} \ln(C_2(t/\tau_0) + C_1), \; v = (4/k)d(-\zeta/k)/dt \quad (12.3)$$

c. Acceleration exponent -2

At $a = -2$, the acceleration is $g = Gt^{-2}$, the time-scale is $\tau_0 = (kv_0)^{-1}$, the acceleration strength is parameterized by $(Gk)$, and the system is transformed to

$$d^2\sigma/dT^2 - A\left[\sigma T^{-2}(Gk) - 2(d\sigma/dT)^2\right] = 0, \; d\sigma/dT - (V/4) = 0 \quad (13)$$

For strong acceleration $Gk \gg 1$ with terms related as $\left|\sigma T^{-2}(Gk)\right| \gg \left|2(d\sigma/dT)^2\right|$, the dynamics is acceleration-driven, and the system is transformed to

$$d^2\sigma/dT^2 - (AGk)\sigma T^{-2} = 0, \; d\sigma/dT - (V/4) = 0 \quad (14.1)$$

The solution of the system is

$$\sigma = C_1 T^{p_1} + C_2 T^{p_2}, \; p_{1,2} = \left(1 \pm \sqrt{1 + 4AGk}\right), \; V = 4[d\sigma/dT] \quad (14.2)$$

In the limiting case of strong acceleration $AGk \gg 1$

$$\sigma = C_1 T^p + C_2 T^{-p}, \; p = \sqrt{4AGk}, \; V = 4[d\sigma/dT] \quad (14.3)$$

and, in the dimensional form, the solution is

$$(-\zeta/k) = C_1(t/\tau_0)^p + C_2(t/\tau_0)^{-p}, \; p = \sqrt{4AGk}, \; v = (4/k)[d(-\zeta/k)/dt] \quad (14.4)$$

For weak acceleration $Gk \ll 1$ with terms related as $|\sigma T^{-2}(Gk)| \gg |2(d\sigma/dT)^2|$, the dynamics is driven by the initial growth-rate, and the system is transformed to

$$d^2\sigma/dT^2 + (2A)(d\sigma/dT)^2 = 0, \; V - 4(d\sigma/dT) = 0 \quad (15.1)$$

The solution of the system in the dimensionless and dimensional forms is

$$\sigma = (2A)^{-1} \ln(C_2 T + C_1), \; V = 4(d\sigma/dT) \quad (15.2)$$

$$(-\zeta/k) = (2A)^{-1} \ln(C_2(t/\tau_0) + C_1), \; v = (4/k)d(-\zeta/k)/dt \quad (15.3)$$

d. Short time intervals

For very short time intervals, $(t-t_0)/t_0 \ll 1$, with $t_0 > \{\tau_G, \tau_0\}$, the governing equations can be linearized. By using the sign function $sgn$, the solution is

$$\zeta - \zeta(t_0) = -(k^2/4)v_0(t-t_0)sgn[v(t_0)/v_0], \; v - v(t_0) = -((Ak/2)v_0^2 + (4A/k)\zeta_0 G t_0^a)(t-t_0) \quad (16)$$

in full consistency with the foregoing results [23,24].

3.2 Effect of initial conditions

Our general solutions are applicable for any sign of $\zeta_0 k$ and $v(t_0)/v_0$. To study qualitatively how the bubbles and spikes are being formed in the early time dynamics [7,8,31], we represent the short time interval solution, for $0 < (t-t_0)/t_0 \ll 1$ with $t_0 \gg \{\tau_G, \tau_0\}$, in the form

$$\zeta - \zeta(t_0) = -(k/4)[(t-t_0)/\tau_0]sgn[v(t_0)/v_0], \quad (17)$$
$$v - v(t_0) = -(A/2)v_0[(t-t_0)/\tau_0] - 4A(\zeta_0/k)(\tau_G k)^{-1}(t_0/\tau_G)^a[(t-t_0)/\tau_G]$$

Two regimes are clearly seen from this expression: the acceleration-driven regime, with $\tau_G/\tau_0 \ll 1$ and $|v_0(\tau_G/\tau_0)| \ll |8(\zeta_0/k)(\tau_G k)^{-1}(t_0/\tau_G)^a|$, and the initial growth-rate-driven regime, with $\tau_0/\tau_G \ll 1$ and $|v_0(\tau_G/\tau_0)| \gg |8(\zeta_0/k)(\tau_G k)^{-1}(t_0/\tau_G)^a|$.

In the acceleration-driven regime, $\tau_G/\tau_0 \ll 1$, the morphology and the velocity of the interface near the tip change with time as follows: For $v(t_0)/v_0 > 0$ and $\zeta_0 k < 0$, the interface becomes more curved and its velocity increases, since $(\zeta - \zeta(t_0))/k < 0, (v - v(t_0))(\tau_G k) > 0$. For $v(t_0)/v_0 > 0$ and $\zeta_0 k > 0$, the interface flattens and the velocity decreases, $(\zeta - \zeta(t_0))/k < 0, (v - v(t_0))(\tau_G k) < 0$. For $v(t_0)/v_0 < 0$ and $\zeta_0 k > 0$, the interface becomes more curved and the velocity magnitude increases, $(\zeta - \zeta(t_0))/k > 0$, $(v - v(t_0))(\tau_G k) < 0$. For $\zeta_0 k < 0$ and $v(t_0)/v_0 < 0$, the interface flattens and the velocity magnitude decreases, $(\zeta - \zeta(t_0))/k > 0$, $(v - v(t_0))(\tau_G k) > 0$. This suggests that the bubbles are formed at the regular points of the interface with $\zeta_0 k < 0$, whereas the spikes are formed at the regular points of the interface with $\zeta_0 k > 0$. In the acceleration-driven dynamics, the positions of the bubbles and spikes are set by the initial morphology of the interface.

In the initial growth-rate driven regime, $\tau_0/\tau_G \ll 1$, the morphology and the velocity of the interface near the tip change with time as follows: For $v(t_0)/v_0 > 0$ and $\zeta_0 k < 0$, the interface becomes more curved and its velocity decreases, since $(\zeta - \zeta(t_0))/k < 0$, $(v - v(t_0))(\tau_G k) < 0$. For $v(t_0)/v_0 > 0$ and $\zeta_0 k > 0$, the interface flattens and velocity decreases, $(\zeta - \zeta(t_0))/k < 0$, $(v - v(t_0))(\tau_G k) < 0$. For $v(t_0)/v_0 < 0$ and $\zeta_0 k > 0$, the interface becomes more curved and velocity magnitude increases, since $(\zeta - \zeta(t_0))/k > 0$, $(v - v(t_0))(\tau_G k) > 0$. For $v(t_0)/v_0 < 0$ and $\zeta_0 k < 0$, the interface flattens and the velocity magnitude increases, $(\zeta - \zeta(t_0))/k > 0$, $(v - v(t_0))(\tau_G k) < 0$. This suggests that the bubbles are formed at the regular points of the interface with $v(t_0)/v_0 > 0$, and the spikes are formed at the regular points of the interface with $v(t_0)/v_0 > 0$. In the initial growth-rate driven dynamics, the positions of the bubbles and spikes are set by the initial velocity field.

3.3 Theory benchmarks

According to our results, for variable acceleration with power-law time-dependence, for the acceleration exponents $a > -2$ and at $a = -2, Gk \gg 1$, the early time dynamics is driven by the acceleration and the positions of the bubbles and spikes are set by the interface morphology, similarly to the case of RTI with constant acceleration. Hence we call the acceleration-driven dynamics as being 'RT-type' [3,4,7,9,10,29,31]. For the acceleration exponents $a < -2$ and at $a = -2, Gk \ll 1$, the early time dynamics is driven by the initial growth-rate and the positions of the bubbles and spikes are set by the initial velocity field. This dynamics is similar to the case of the Richtmyer-Meshkov instability, where the growth of the interface perturbation is due to impulsive acceleration by the shock; in a broad parameter regime the initial growth-rate is constant and the associated motion is nearly incompressible. Hence we call the initial-growth-rate-driven dynamics as being 'RM-type' [4,5,7,10,16,29,31].

Note that in the RTI and RMI in supernova blasts and in ICF environments, typical values of the acceleration exponents are $a > -2$ [17,18,24,26] According to our results, while in these cases the acceleration is usually induced by unsteady shocks, the early time dynamics of the unstable flow is RT-type: It is defined by the acceleration parameters, with the positions of bubbles and spikes set by the initial morphology of the interface [3,4,7,9,10,29,31].

Our results identify theory benchmarks for experiments and simulations, which, to our knowledge, have not been discussed before [7-24]. Specifically, by implementing in experiments and simulations an acceleration with exponents $a > -2$ one can study RT-type dynamics, and observe super-exponential growth of the interface perturbations for $a > 0$ and sub-exponential growth for $-2 < a < 0$. By implementing in experiments and simulations an accelerations with exponents $a < -2$ one can study RM-type dynamics, and observe the growth of the interface perturbations, which is set by the initial growth-rate and is independent of the exponent. By implementing an acceleration with exponent $a = -2$, one can further observe the effect of the acceleration strength on the dynamics.

In addition to quantitative study of dependence of the instability growth-rate on the acceleration parameters and the initial growth-rate, one can investigate the dependence of the unstable dynamics on the initial conditions. Particularly, for RT-type dynamics, one can observe that the formation of bubbles and spikes structure is prescribed by the morphology of the initially perturbed interface. This result is in excellent agreement with experiments [9-16,29]. For RM-type dynamics, one can further observe that the process of formation of the structure of bubbles and spikes is prescribed by the initial velocity field at the interface. For some initial conditions, it may lead to the so-called 'phase reversal', with bubbles (spikes) turning to spikes (bubbles), in excellent agreement with experiments and simulations [12,29,31].

The other important diagnostic parameter found by our theory is the qualitative velocity field [23,24]. According to our results in RT- and RM-type dynamics the non-equilibrium velocity field is characterized by effectively no motion of the fluids away from the interface, intensive motion of the fluids in a vicinity of the interface, and the production of shear-driven vortical structures at small scales at the interface [7,8,23,24,30]. This result excellently agrees with experiments and simulations [9-16]. Note that for a shock-induced acceleration, the non-equilibrium velocity is referred to the velocity in a frame of reference moving with the velocity of the background motion [5,6,12,10,16]. This is because for a spatially extended periodic flow the post-shock dynamics is a superposition of two motions – the background motions of the fluids and the interface in the transmitted shock direction, and the growth of the interface perturbations due to impulsive acceleration by the shock. The velocity scale of the background motion is substantially greater than the initial growth-rate; for strong shock the former is usually super-sonic, whereas the latter is sub-sonic [16].

Our analysis can be applied for three-dimensional flows with other symmetries and two-dimensional flow [7,8,24]. Particularly, for three-dimensional highly symmetric flows, the dynamics is universal, expect for the difference in the wavevector value for a given $\lambda$ [7,8,30]. This universality is due to a nearly isotropic character of the dynamics in the plane normal to the acceleration [7,8,30].

**Section 4 – Discussion**

We have studied the long-standing problem of the early-time dynamics of Rayleigh-Taylor instability with time-variable acceleration and for a spatially extended three-dimensional flow periodic in the plane normal to the acceleration with symmetry group p6mm, Eqs.(1-17). For the acceleration with power-law time-dependence, $g = Gt^a$, by employing group theory and scaling analysis, we have explicitly found the instability growth-rate in a broad range of the acceleration parameters, have investigated the dependence of the dynamics on the initial conditions, and have elaborated theory benchmarks. Two distinct sub-regimes of the early time dynamics, depending on the acceleration exponent, have been discovered. To our knowledge, these findings have not been discussed before [1-31].

Particularly, we have found that for exponents $a > -2$, the time-scale is $\tau_G = (kG)^{-1/(a+2)}$, and the dynamics is of the acceleration-driven RT-type, Eqs.(11). For exponents $a < -2$, the time-scale is $\tau_0 = (kv_0)^{-1}$, and the dynamics is of the initial growth-rate driven RM-type, Eqs.(12). At the exponent $a = -2$, the time-scale is $\tau_0 = (kv_0)^{-1}$, and the dynamics changes its character from RT to RM-type with decrease of the acceleration strength $Gk$, Eqs.(14,15) For short time intervals, the solution depends linearly on the time interval, Eqs.(16). The formation of the structure of bubbles and spikes is prescribed by the initial morphology of the interface for RT-type dynamics, and by the initial velocity field for RM-type dynamics, Eqs.(17).

Our results are in excellent agreement with available experiments and simulations [9-17,24,29], and elaborate theory benchmarks for future experiments and simulations. These include the dependence of the instability growth-rate on the acceleration parameters and the initial growth-rate, the dependence of the process of formation of the structure of bubbles and spikes on the initial morphology of the interface and the initial velocity filed, along with qualitative properties of non-equilibrium flow fields, including the intense motions of the fluids near the interface and effectively no motion of the fluids away from the interface.

Our analysis can be extended to study advanced stages of RTI and RT mixing [24], and to systematically account for the properties of non-ideal fluids, including the effects of compressibility, surface tension, and viscosity [10-20,27,28], to be done in future research.

**Acknowledgements:**
The authors thank the University of Western Australia (AUS); the National Science Foundation (USA).